\documentclass[
aps,a4paper,
reprint,
twocolumn,
showpacs,showkeys,
superscriptaddress,
floatfix,
preprintnumbers,
amsmath,amssymb
]{revtex4-1}
\usepackage[utf8]{inputenc}
\usepackage{xspace}
\usepackage{graphicx}
\usepackage{epstopdf}
\usepackage{color,xcolor}
\usepackage{amsmath,amsfonts,amssymb}
\usepackage{bm}
\usepackage{upgreek}
\usepackage{nicefrac}
\usepackage{siunitx}
\DeclareSIUnit\atper{{at.\,\percent}}
\usepackage[colorlinks=true]{hyperref}
\newcommand{\etal}{\textit{et~al.}\xspace}

\newcommand{\sfmo}{{Sr$_2$FeMoO$_6$}\xspace}
\newcommand{\vO}  {\ensuremath{\mathrm{V}_\mathrm{O}}\xspace}
\newcommand{\vOxy}{\ensuremath{\mathrm{V}_{\mathrm{O}_{xy}}}\xspace}
\newcommand{\vOz} {\ensuremath{\mathrm{V}_{\mathrm{O}_z}}\xspace}
\newcommand{\vSr} {\ensuremath\text{V}_\text{Sr}\xspace}
\newcommand{\vFe} {\ensuremath\text{V}_\text{Fe\xspace}}
\newcommand{\vMo} {\ensuremath\text{V}_\text{Mo}\xspace}
\newcommand{\muB}{\mu_\text{B}}      

\newcommand{\due}{e_\text{g}^{\uparrow}}

\newcommand{\dut}{t_\text{2g}^{\uparrow}}

\newcommand{\Fref}[2][]{Fig.~\ref{#2}\textcolor{black}{#1}} 

\newcommand{\Fsref}[2][]{Figs.~\ref{#2}\textcolor{black}{#1}} 

\newcommand{\Tref}[1]{Tab.~\ref{#1}\xspace} 

\newcommand{\cref}[1]{chapter~\ref{#1}\xspace}
\newcommand{\Cref}[1]{Chapter~\ref{#1}\xspace} 

\usepackage{pdfpages}
\usepackage{etoolbox} 

\makeatletter
\patchcmd{\@outputpage@head}{\@ifx{\LS@rot\@undefined}{}{\LS@rot}}{}{}{}
\makeatother
\begin{document}
\author{Waheed A. Adeagbo}
\email{waheed.adeagbo@physik.uni-halle.de}
\affiliation{Institute of Physics, Martin Luther University Halle-Wittenberg, Von-Seckendorff-Platz 1, 06120 Halle, Germany}

\author{Martin Hoffmann}
\email{martin.hoffmann@jku.at}
\affiliation{Institute for Theoretical Physics, Johannes Kepler University Linz, Altenberger Stra\ss{}e 69, 4040 Linz, Austria}

\author{Arthur Ernst}
\affiliation{Institute for Theoretical Physics, Johannes Kepler University Linz, Altenberger Stra\ss{}e 69, 4040 Linz, Austria}
\affiliation{Max Planck Institute of Microstructure Physics, Weinberg 2, 06120 Halle, Germany}

\author{Wolfram Hergert}
\affiliation{Institute of Physics, Martin Luther University Halle-Wittenberg, Von-Seckendorff-Platz 1, 06120 Halle, Germany}

\author{Minnamari Saloaro}
\affiliation{Wihuri Physical Laboratory, Department of Physics and Astronomy, University of Turku, FI-20014 Turku, Finland}

\author{Petriina Paturi}
\affiliation{Wihuri Physical Laboratory, Department of Physics and Astronomy, University of Turku, FI-20014 Turku, Finland}

\author{Kalevi Kokko}
\affiliation{Department of Physics and Astronomy, University of Turku, FIN-20014 Turku, Finland}
\affiliation{Turku University Centre for Materials and Surfaces (MatSurf), Turku, Finland}

\title{Tuning the probability of defect formation via
  substrate strains in \texorpdfstring{\sfmo}{Sr2FeMoO6} films} 

\date{\today}

\begin{abstract}
Since oxide materials like \sfmo are usually applied as thin films,
we studied the effect of biaxial strain, resulting 
from the substrate, on the electronic and 
magnetic properties and, in particular, on the formation energy
of point defects. From our first-principles 
calculations, we determined that the probability of forming 
point defects -- like vacancies or substitutions  -- 
in \sfmo could be adjusted by choosing a proper substrate.
For example, the amount of anti-site disorder  
can be reduced with compressive strain in order to obtain purer \sfmo as needed for 
spintronic applications, while the formation 
of oxygen vacancies is more likely for tensile strain, which 
improves the functionality of \sfmo as a basis material
of solid oxide fuel cells. 
In addition, we were also be able to include the 
oxygen partial pressure in our study by using 
its thermodynamic connection with the chemical potential. Strontium
vacancies become for example
more likely than oxygen vacancies
at a pressure of \SI{1}{\bar}.
Hence, this degree of freedom might offer in general
another potential method for defect engineering in oxides
besides, e.g.,
experimental growth conditions like temperature or gas pressure.
\end{abstract}

\maketitle

\section{Introduction}

Lattice imperfections like point defects can be crucial for the functionality 
of novel materials. They can destroy desired properties but might also improve
or even just allow new functionalities. A good example is the double 
perovskite material Sr$_2$Fe$_{1+x}$Mo$_{1-x}$O$_{6-\delta}$, where $x$
symbolizes non-stoichiometry at the $B$/$B'$ site and $\delta$ oxygen-deficiency
(general formula $A_2BB'$O$_6$). It
is a versatile material, which is throughout literature 
considered for several potential applications. 

For example, the half-metallic character of \sfmo (SFMO),
which is experimentally observed as promising high spin polarization
\cite{Panguluri2009apl},
is a highly desired property in order to access
spintronics
\cite{Bibes2003_10.1063/1.1612902,
  Nag2017_10.1007/s12648-017-0986-2}.
Besides, SFMO shows a low-field magnetoresistance response,
a magnetic transition above ambient temperatures
\cite{Fontcuberta2005jmmm},
and a high magnetic moment per functional unit in a range of
$2.2\muB/\si{{f.u.}}$ to $3.9\muB/\si{{f.u.}}$
\cite{Yin1999, Manako1999apl, Tomioka2000prb, DiTrolio2006jap}.
Nevertheless, a successful utilization of SFMO is still
problematic, since the theoretical predicted half-metallicity
\cite{Kobayashi1998n}, which is one requirement of its
high spin polarization, is usually
diminished or even lost in thin film samples of \sfmo
\cite{Jalili2009prb,
  Saloaro2013_10.1051/epjconf/20134015012,
  Angervo2015_10.1016/j.phpro.2015.12.170,
  Saloaro2016_10.1021/acsami.6b04132}.
These deterioration effect was attributed 
to doping, defects, grain boundaries, or parasitic phases,
so that, e.g., the saturation magnetization becomes strongly 
reduced compared with its theoretically expected value. 
The latter should be $4\muB/\si{{f.u.}}$
arising from the spin quantum number of $S = 5/2$ of the 
Fe$^{3+}$ ions and the antiparallel coupled electrons with $S = 1/2$
from the Mo$^{5+}$ ions \cite{Kobayashi1998n}.
It was verified in many theoretical studies, which 
shed light upon the microscopic origins of the magnetic 
properties of SFMO and their alterations
\cite{Ogale1999apl, Navarro2001jpcm,
  Chan2003_10.1021/cm020773h,
  Colis2005jap_10.1063/1.1997286, 
  Kahoul2008_10.1063/1.3043586, Mishra2010cm,
  MunozGarcia2011_10.1021/cm201799c}.
While oxygen vacancies --
a common observation in transition metal oxides -- seem not to interfere 
with the spin polarization of SFMO
\cite{Stoeffler2005jmmm, 
  MunozGarcia2011_10.1021/cm201799c,
  Zhang2015jac_1,
  Saloaro2016_10.1021/acsami.6b04132},
the stoichiometry preserving exchange of $B$ and $B'$ atoms, so-called 
antisite disorder (ASD), can
strongly reduce the saturation magnetization, 
the spin polarization \cite{Stoeffler2005jmmm,Zhang2015jac_1,Reyes2016jpcc},
the magneto-transport properties
\cite{Toepfer2009_10.1063/1.3065969, Deniz2017jap_10.1063/1.4973878}
or might even influence 
long range magnetic ordering
\cite{Serrate2007jpcm,Erten2013prb}. Hence, 
a synthesis of \sfmo samples is rather difficult
because of a non-negligible concentration of defects. Therefore,
there is 
a strong desire to control the amount of defects in \sfmo samples
\cite{Saloaro2013_10.1051/epjconf/20134015012,
  Saloaro2016_10.1021/acsami.6b04132}.

On the other hand, the easy formation of oxygen vacancies 
in \sfmo makes it a good candidate as mixed ionic electronic 
conductor in solid oxide fuel cells (SOFC) 
\cite{MunozGarcia2011_10.1021/cm201799c,
  Gomez2016_10.1016/j.rser.2016.03.005}.
Experimental studies investigate its potential as anodes
with a non-stoichiometric ratio between
Fe and Mo
\cite{Li2016_10.1016/j.cattod.2015.04.025,
  Miao2016_10.1016/j.ijhydene.2015.12.045},
while theoretical studies calculate the oxygen diffusion
\cite{MunozGarcia2014_10.1021/ar4003174} or 
the formation energy of oxygen vacancies in bulk \sfmo
\cite{MunozGarcia2011_10.1021/cm201799c}.

Both applications utilize thin films of SFMO 
\cite{Kovalev2014_10.1021/am5052125,
  Saloaro2016_10.1021/acsami.6b04132}, 
where the substrate might cause epitaxial biaxial strain
-- another factor deteriorating the desired 
properties of the SFMO film \cite{Fix2005jap}. 
Although very large biaxial strains could theoretically induce a spin transition  
in bulk SFMO \cite{Lu2014ssc},
our recent study revealed not only that a compressive biaxial strain $\sim\SI{1}{\percent}$
at SrTiO$_3$ substrates does 
not show any significant direct impact on the magnetic properties of SFMO
\cite{Saloaro2016_10.1021/acsami.6b04132}, but also that 
the amount of ASD and oxygen vacancies is reduced
compared to the observed defect concentrations in the SFMO bulk target
from which the films were grown
\cite{Saloaro2016_10.1021/acsami.6b04132}.

This increase of defect formation probabilities was an interesting experimental
finding. 
Hence, we want to study the role of biaxial strain in SFMO
further in more detail. In particular, oxygen vacancies
are one way for various oxides to compensate tensile strain, 
since the observed chemical expansion can accommodate parts of the strain
\cite{Aschauer2013prb,Aschauer2015prb}. 
It could mean therefore potentially room 
temperature magnetism for SFMO thin films due to tensile 
biaxial strain because oxygen vacancies
were shown to increase the Curie temperature in SFMO as well
\cite{Hoffmann2015arxiv_SFMO}.

However, other point defect than oxygen vacancies
were only rarely discussed in literature on strained oxide materials. 
We carried out density-functional theory (DFT) calculations in order
to study systematically the structural and magnetic properties of SFMO
including various point defects under the influence of biaxial strain,
in particular, ASD and other vacancies.
We observe strong variations of formation energies up to \SI{1}{\eV}
depending on the kind of considered defect allowing for
potential defect engineering with the decision for a particular substrate. 

\section{Technical details}

Our calculations are performed using the 
Vienna {\textit{ab initio}} simulation package (VASP) \cite{kresse1993prb,kresse1996cms}.
While we present here only setup information,
which are deviating from a standard VASP setup used, e.g., in previous 
theoretical studies on SFMO
\cite{MunozGarcia2011_10.1021/cm201799c,Zhang2015jac_1}, 
we provide further numerical details and parameters in the Supplemental Material 
\footnote{See
  Supplemental Material below for
  details about the computational setup, the electronic structure of SFMO,
  and the calculation of defect formation energies.}.
\begin{figure}
  \begin{center}
    \includegraphics[width=246pt]{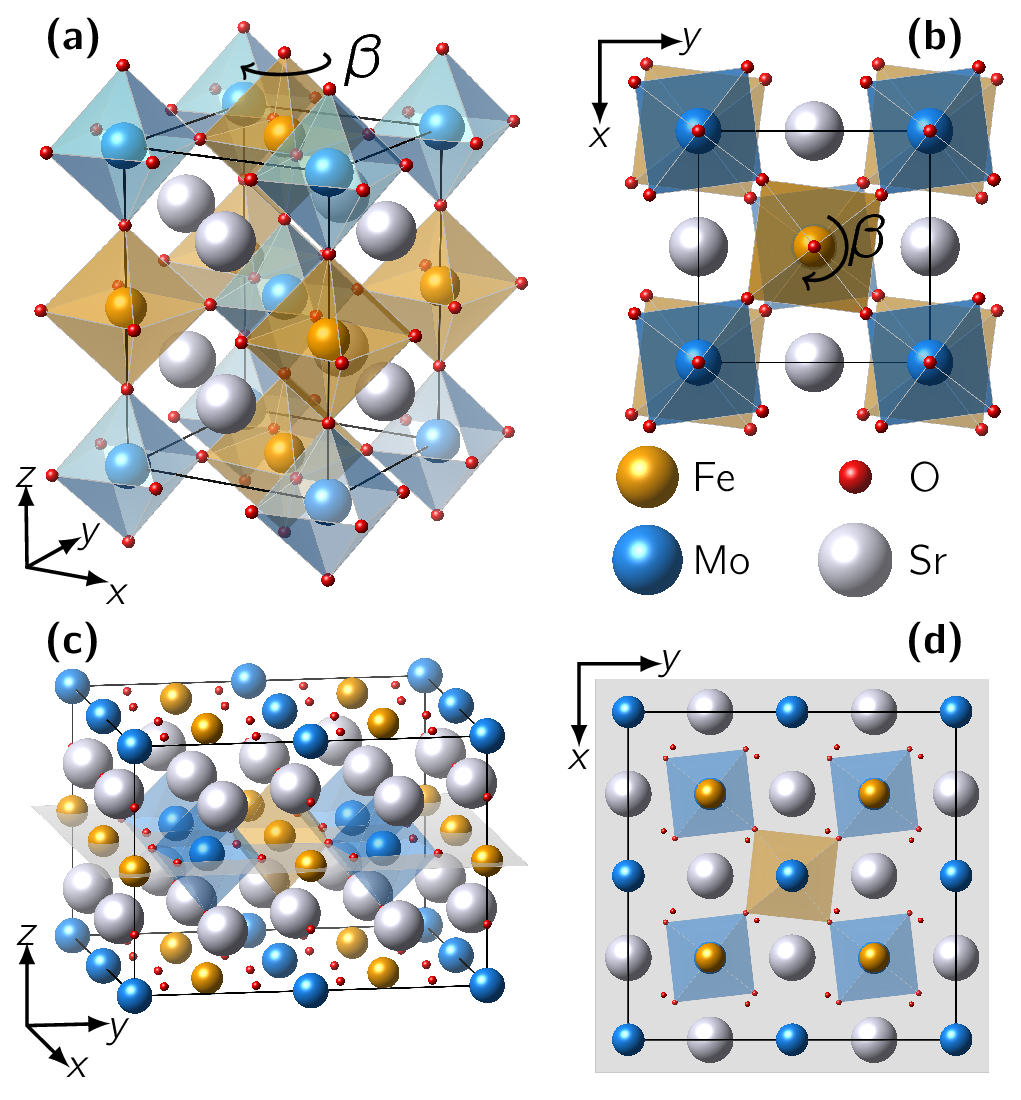}
  \end{center}
  \caption{(Color online) Considered lattice structures of {\sfmo}:
    (a) Tetragonal unit cell with   $I4/m$  symmetry visualizing
    the octahedral environment of Fe and Mo formed by the oxygen
    atoms with colored polyhedra. (b) Top view showing the 
    octahedra rotation denoted 
    as $\beta$ and counted in clockwise direction for the Fe ion octahedron.
    (c) Side view of the $2\times2\times1$ supercell with
    \SI{8}{{f.u.}} used for the  defect calculations. 
    (d) Top view of the supercell in (c) from the gray $xy$ 
    layer downwards.
  }
  \label{fig:structure} 
\end{figure}

We used in our calculations the tetragonal unit cell containing two 
functional units (\si{{f.u.}}) of \sfmo
(\Fsref{fig:structure}a and \ref{fig:structure}b)
and a supercell comprising \SI{8}{{f.u.}} of {\sfmo} -- in total
80 atoms (\Fsref{fig:structure}c and \ref{fig:structure}d).
Both structures stabilize after numerical relaxation in 
the space group $I4/m$, which is in agreement with experimental
observations and shows oxygen octahedra, which rotate
  only slighly in the $x-y$ plane and remain static otherwise
(\Fref{fig:structure}). The Fe octahedra are rotated clockwise
  by $\beta=\SI{7.43}{\degree}$ while the Mo octahedra rotate counter clockwise.

Former theoretical studies \cite{Mishra2010cm,
  MunozGarcia2011_10.1021/cm201799c,Lu2014ssc,Zhang2015jac_1} discuss
correlation corrections GGA$+U$ as the crucial factor in order to
obtain a total magnetic moment of $4.0\muB/\si{{f.u.}}$, but we will
show below that this is also achieved with the fixed spin moment (FSM)
method
\cite{Schwarz1984_10.1088/0305-4608/14/7/008,Marcus1988_10.1063/1.340544}
and a GGA functional alone. Besides, the band gap in the spin up
channel obtained within the GGA$+U$
approach is too wide compared with experimental measured values
of \SI{0.5}{\eV}, \SI{1.3}{\eV} \cite{Note1,Tomioka2000prb,
  Saitoh2002_10.1103/PhysRevB.66.035112}.

The FSM method allows instead to treat the total energy as an
explicit function of the magnetization. It was in particular 
important for regions of large compressive or tensile strain
where the total energy landscape can be 
strongly varied and where it becomes more difficult 
to ascertain the exact magnetic
moments of individual ions.

\section{Ground State Properties with Fixed Spin Moment Method} 

At first, we studied defect-free SFMO in order to demonstrate that 
the FSM approach is able to reproduce the correct structural 
ground state properties of SFMO. 
Therefore, as a benchmark calculation,
we imposed with FSM a magnetic moment on the defect-free tetragonal SFMO unit cell and 
calculated the total energy in order to search for the
most stable spin configuration. 
The configuration with a net magnetic moment (MM) of $m_\mathrm{gs}= 4.0\muB/\si{{f.u.}}$
had finally the lowest
total energy and is, therefore, the ground state at \SI{0}{\kelvin} for defect-free unstrained SFMO.
We have to note that this ground state shows directly the half-metallic 
density of states (DOS) with only spin down states at the 
Fermi energy. The additional application of correlation corrections 
within GGA$+U$ ($U=\SI{4}{\eV}$ on Fe-3$d$ orbitals) on top of FSM leads to
a wider half-metallic band gap (from \SI{1.3}{\eV} to \SI{2.4}{\eV}), 
while the magnetic moment at the Fe site
increases from $m_\mathrm{Fe}=3.6\muB/\si{{f.u.}}$ to $m_\mathrm{Fe}=3.8\muB/\si{{f.u.}}$

\begin{table}
  \renewcommand{\arraystretch}{1.}
  \caption{Structural properties of defect-free, unstrained \sfmo with FSM:
    bulk modulus $B_0$, Gr\"uneisen constant $B_0'$, 
    and the lattice parameters. The units are given in brackets, while $B_0'$ is 
    dimensionless. Experimental results for polycrystalline \sfmo samples
    obtained at room temperature
    vary in grain size from \SI{50}{\nm} or \SI{100}{\nm} \cite{E2006jms}
    to $\sim\SI{2}{\micro\meter}$ \cite{Zhao2002jap}.}
  \begin{center}
    \begin{tabular*}
      {\columnwidth}
      {@{\extracolsep{\fill}}l  c c   ccc }
      \hline\hline
      &  GGA     & GGA$+U$ & \multicolumn{3}{c}{Experiment \cite{Zhao2002jap,E2006jms} }\\\hline
      $B_{0}$  (\si{\GPa})    &  147.42      & 152.80   &  \num{266\pm3} & \num{284\pm6} & \num{331\pm12} \\ 
      $B'_0$     &  4.68        & 4.44     &  4.0 & 4.0 & 4.0  \\  
      $a$ (\si{\angstrom})    & 5.5522      & 5.6378   & 5.5703 & 5.5791 & 5.5703 \\  
      $c$ (\si{\angstrom})    & 7.9013      & 7.9731   & 7.7879 & 7.8698 & 7.8832 \\  
      grain size (\si{\nm})   & &  & $\sim2000$ & 100 & 50 \\
      \hline \hline
    \end{tabular*}
    \label{tab:B}
  \end{center}
\end{table}  

Indeed, also the corresponding lattice parameters of 
defect-free SFMO at the ground state
($a=b=\SI{5.55}{\angstrom}$ and $c=\SI{7.90}{\angstrom}$)
agree very well with various experimental results
\cite{Zhao2002jap,E2006jms, 
  Saloaro2013_10.1051/epjconf/20134015012,Taylor2015cc}
for the FSM method alone (\Tref{tab:B}).
In fact, they are the same as the lattice parameters measured at
\SI{70}{\kelvin} 
\cite{Chmaissem2000_10.1103/PhysRevB.62.14197}
($a=\SI{5.5521}{\angstrom}$ and $c=\SI{7.9013}{\angstrom}$).
This agreement is much better than what can be achieved 
by the GGA$+U$ method obtained earlier
\cite{Mishra2010cm,MunozGarcia2011_10.1021/cm201799c, Lu2014ssc}
(see also our results in \Tref{tab:B}).

We calculated also the bulk modulus $B_0$ via the hydrostatic pressure
but available experimental data \cite{Zhao2002jap,E2006jms}
is much larger than the numerical value (\Tref{tab:B}).
However, we note that these experimental
bulk moduli were determined with polycrystalline samples and the bulk modulus decreases strongly
by \SI{65}{\GPa} with increasing the  grain size from \SI{50}{\nm} to \SI{2}{\micro\meter}.
We can interpret our theoretical {\textit{ab initio}} calculation
as a kind of limit to the infinite large grain size,
which could then validate the bulk modulus (\SI{147}{\GPa} to \SI{153}{\GPa})
obtained in our and former calculations \cite{MunozGarcia2011_10.1021/cm201799c}.
Applying additional correlation corrections (GGA$+U$) shows here
only small variations $\sim\SI{5}{\GPa}$ (\Tref{tab:B}).

\begin{table}
  \renewcommand{\arraystretch}{1.}
  \caption{Single crystal elastic constants
    $c_{ij}$ of defect-free, unstrained \sfmo with FSM
    written in Voigt notation and in GPa.}
  \begin{center} 
    \begin{tabular*}
      {\columnwidth}
      {@{\extracolsep{\fill}}l *{6}{c} }
      \hline\hline
      & $c_{11}$  & $c_{12}$ & $c_{13}$ &  $c_{33}$ &  $c_{44}$ & $c_{66}$ \\\hline
      GGA     &  230.32 & 77.70 & 108.93 & 244.42 & 45.48 & 63.95 \\
      GGA$+U$ &  245.54 & 82.97 & 109.00 & 267.97 & 48.489& 76.61 \\
      \hline \hline
    \end{tabular*}
    \label{tab:cjs}
  \end{center}
\end{table} 

Finally, we determined the elastic tensor of the
tetragonal structure by six independent
single crystal elastic constants: $c_{11}$, $c_{12}$,  $c_{13}$,   $c_{33}$,  $c_{44}$ and  $c_{66}$
(\Tref{tab:cjs}). Experimentally determined elastic constants 
were unfortunately not available,
but theoretical values for other double perovskites
do not deviate far from our results (\Tref{tab:cjs}):
Ba$_2$MgWO$_6$ has $c_{11}=\SI{256.2}{\GPa}$, 
$c_{12}=\SI{85.8}{\giga\pascal}$, 
$c_{44}=\SI{95.3}{\giga\pascal}$, and 
$B_0=\SI{143.1}{\giga\pascal}$ 
\cite{Shi2013_10.1140/epjb/e2012-30584-1}; or Sr$_2$CaOsO$_6$ has
$c_{11}=\SI{331.19}{\giga\pascal}$, 
$c_{12}=\SI{77.47}{\giga\pascal}$, 
$c_{44}=\SI{63.35}{\giga\pascal}$, and 
$B_0=\SI{162.04}{\giga\pascal}$
\cite{Faizan2016_10.1007/s12034-016-1288-6}.
For GGA$+U$, the $c_{ij}$ of SFMO, except of $c_{66}$, vary by less than
$\leq\SI{10}{\percent}$ (\Tref{tab:cjs}).

We conclude that 
the GGA and FSM method are indeed a valid tool to calculate the
ground state properties for this particular magnetic material,
while the additional usage of correlation corrections 
does not influence the ground state properties strongly or 
even describes them wrongly. 
Hence, we applied the FSM in all further calculations including
biaxial strain and/or defects:
(a) We set a total magnetic moment for the unit cell and calculated the
structural relaxation.
(b) For a set of varying magnetic moments, we defined the structure with the
lowest total energy as the most stable structural and magnetic configuration with 
total magnetic moment $m_\mathrm{gs}$.
If needed for reasons of comparison, we will mention explicitly the use of correlation
corrections (GGA$+U$).

\section{Variations in bulk SFMO with Biaxial Strain}

\begin{figure}
  \begin{center}
    \includegraphics[width=246pt]{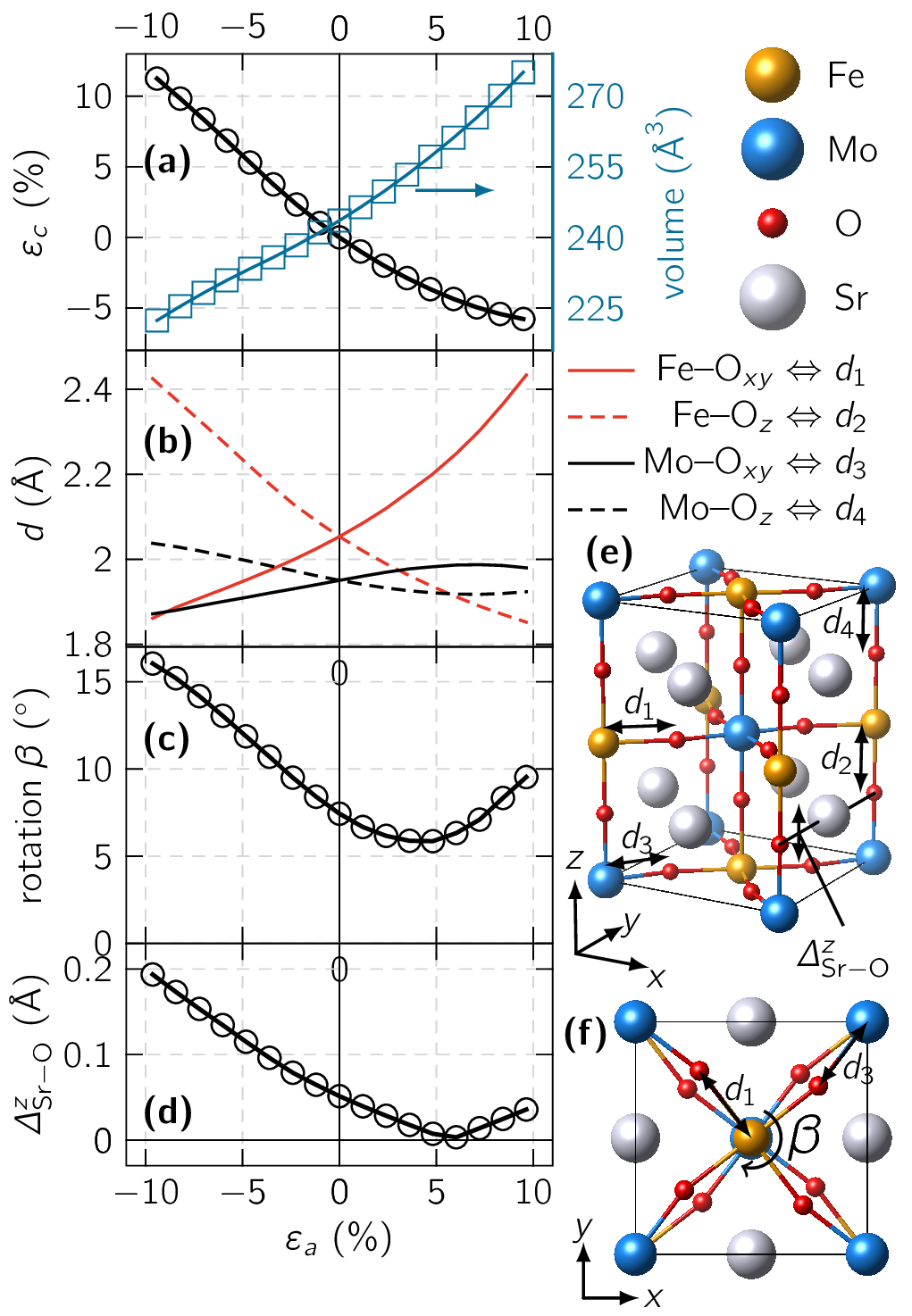}
  \end{center}
  \caption{ (Color online) Structural variation with biaxial strain
    in the lattice constants $a=b$.
    (a) Variation of lattice parameter $c$ and the volume of the tetragonal cell.
    (b) The in-plane ($xy$) and out-of-plane $z$ bond lengths
    Mo-O  and Fe-O.
    (c) Octahedra rotation angle $\beta$, visualized in (f).
    (d) Out-of-plane distances between Sr and O.
    (e)-(f) Visualization of distances and angles in the tetragonal
    unit cell. }
  \label{fig:relax}
\end{figure}

In our previous study \cite{Saloaro2016_10.1021/acsami.6b04132},
only calculations at selected biaxial strain values were considered. 
Here  in the current work, we continued with a comprehensive study 
of SFMO for a larger range of biaxial 
strain going from \SIrange{-10}{10}{\percent}. The structural 
and magnetic configuration at zero strain is represented 
by the tetragonal ground-state
structure (\Fref{fig:structure}a) as described above
and the biaxial strain was applied by fixing the in-plane
lattice constants ($a=b$) and relaxing the out-of-plane
lattice constants ($c$) and the internal parameters (\Fref{fig:relax}).
Although a similar study was already published by Lu \etal \cite{Lu2014ssc},
we present the variations of the structural and magnetic properties of SFMO
with biaxial strain as the basis for the later discussion of the formation energy of point defects.
In addition, we used a different numerical treatment -- namely the FSM method only -- while 
Lu \etal \cite{Lu2014ssc} used the GGA$+U$ method
with $U=\SI{4}{\eV}$.
  
\begin{figure*}
  \begin{center}
    \includegraphics[width=510pt]{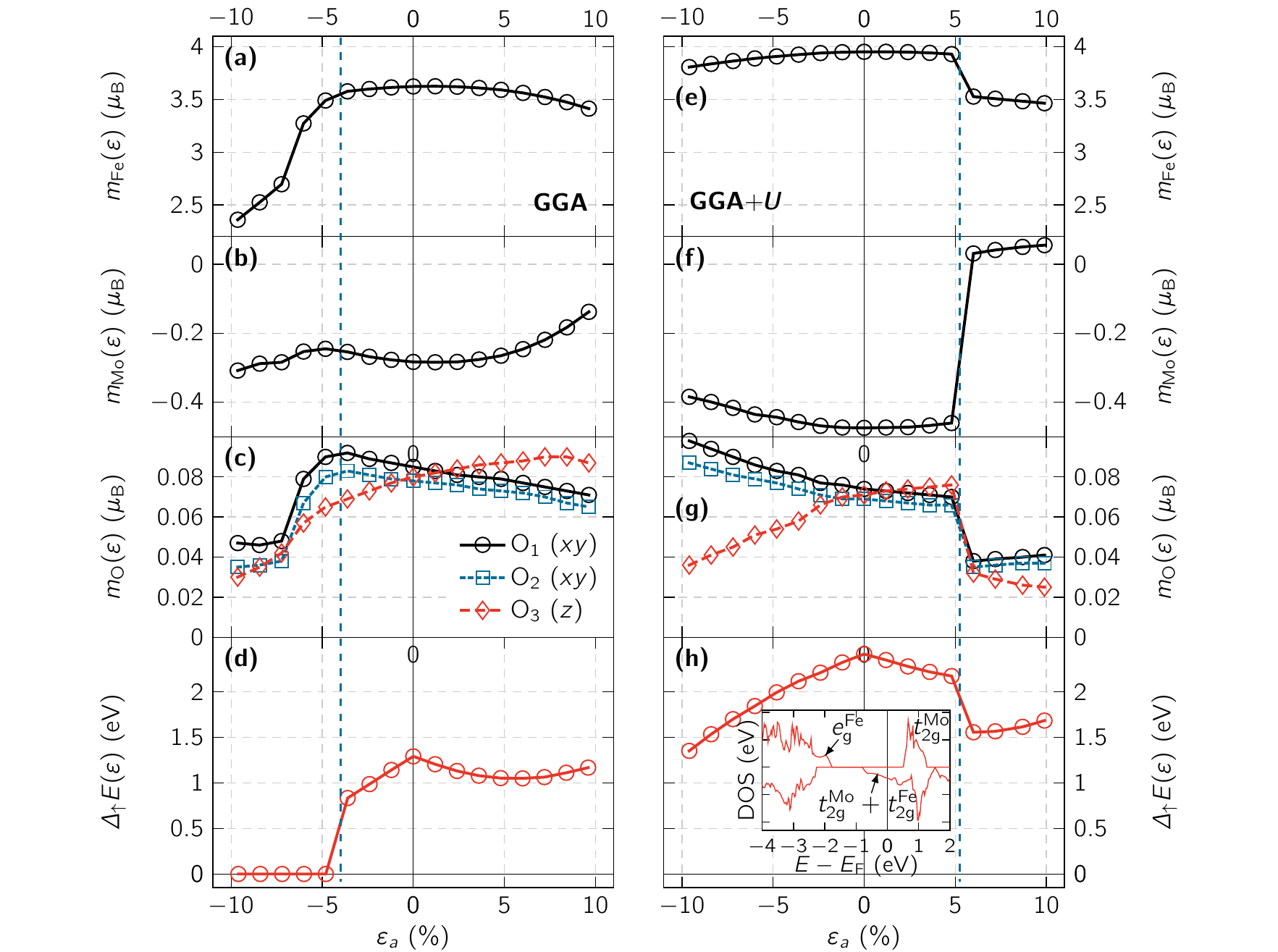}
  \end{center}
  \caption{ (Color online) Variations of magnetic and electronic properties with biaxial strain 
    for GGA and GGA$+U$. The local magnetic moments for (a), (e) Fe ions, 
    (b), (f) Mo ions, and
    (c), (g) O ions. 
    (d), (h) Band gap in the majority spin channel
    calculated from the density of states (DOS).
    The inset shows the DOS for unstrained defect-free SFMO. 
    The vertical dashed line at \SI{-4}{\percent} on the left hand side marks the spin transition from
    half-metallic high-spin state to metallic low-spin state. The vertical dashed line at \SI{5}{\percent}
    on the right hand side marks the transition
    from ferrimagnetic (FIM) to ferromagnetic (FM) state.}
  \label{fig:strain_magnetism} 
\end{figure*}

At each applied biaxial strain, we ensured 
the true ground-state properties using the FSM approach.
We characterized the corresponding structural variation
by analyzing the volume, Fe-O and Mo-O bond lengths, 
the different height of Sr and O ions, 
and the octahedra rotation angles with biaxial strain. 
As observed already for other oxides \cite{Aschauer2013prb,Aschauer2015prb},
SFMO does not follow the Poisson ratio but rather its volume becomes
reduced (increased) for compressive (tensile) biaxial strain (\Fref{fig:relax}a).
The in-plane ($xy$) and out-of-plane ($z$) Fe-O bond length 
(red marks in \Fref{fig:relax}b, distances $d_1$ and $d_2$ in \Fsref{fig:relax}e and  \ref{fig:relax}f),
vary considerably more than the Mo-O bond lengths (black marks in \Fref{fig:relax}b, 
distances $d_3$ and $d_4$ in \Fsref{fig:relax}e and  \ref{fig:relax}f), while 
their values of \SI{2.053}{\angstrom} and \SI{1.951}{\angstrom}, respectively,
for unstrained SFMO
agree well with experimental values \cite{Taylor2015cc}. The bond lengths follow
the natural tendency, the in-plane contributions become smaller with compressive biaxial
strain and elongated for tensile biaxial strain, while it is vice versa for the out-of-plane
bond lengths.
The oxygen ions hybridize stronger
  with the $4d$ states of Mo than with the $3d$ states of Fe.
Already Solovyev \cite{Solovyev2002prb} found that a smaller
Mo-O bond lenght should be the natural state of SFMO.
The bond lengths are also connected with 
the rotation angles of the oxygen octahedra (angle $\beta$ in 
\Fref{fig:relax}f). The rotation becomes larger under compressive biaxial strain
(\Fref{fig:relax}c). Hence, the additional octahedra rotation is another
mechanism in order to compensate the in-plane biaxial strain \cite{Aschauer2013prb}.
In $z$ direction, the elongation caused by the compressive in-plane strain
causes the oxygen ions to hybridize even stronger with Mo, while the 
Fe-O bond becomes much larger (\Fref{fig:relax}b). The 
octahedra experience even stronger breathing distortions than 
in the ground state \cite{Solovyev2002prb}.
The other oxygen ions in the $xy$ plane of Sr-O  avoid in a similar
way the compressive strain by
relaxing out of that plane (\Fref{fig:relax}e). Interestingly, the oxygen
ions do not relax closer to the Fe ions but form an ideal plane with Sr and
relax out of the plane again for larger tensile strains. 

These internal structural changes influence of course
magnetic and electronic properties as it was observed already
earlier \cite{Lu2014ssc}. Using only the FSM methods, we obtain 
only small variations 
for the experimental relevant range of \SI{\pm4}{\percent} biaxial strains.
The Fe ions keep their high spin state with a maximal
local magnetic moment of $3.63\muB/\si{{f.u.}}$  
at zero strain (unstrained SFMO), which is only slightly diminished 
for both strains -- compressive or tensile -- in agreement with
the conclusions of Lu \etal \cite{Lu2014ssc} (\Fref{fig:strain_magnetism}a).
The same holds true 
for the antiparallelly aligned magnetic moments at the Mo ions
-- the absolute value is slightly reduced (\Fref{fig:strain_magnetism}b).
The local magnetic moments at the oxygen ions follow the opposite tendency of 
their bond lengths with the Fe and Mo ions (\Fref{fig:relax}b).
The local moment increases, if the bond length is reduced, and 
decreases, if the bond length is elongated (\Fref{fig:strain_magnetism}c).

The situation changes for compressive strains, which
 are larger than \SI{-4}{\percent}. Here, we obtained a transition
to a completely metallic state because the band gap in the majority
spin channel $\Delta_\uparrow E$ closes (marked with a vertical dashed
line in \Fref{fig:strain_magnetism}d). With the compressive biaxial
strain, the Fe $\due$ states are almost linearly shifted towards the
Fermi energy $E_\mathrm{F}$ (see inset in \Fref{fig:strain_magnetism}h
and \cite{Note1}).
At that strain value when the Fe $\due$ state ``hits'' the Fermi energy,
  a kind of ``jump'' seems to appear. This only follows from the fact
that
$\Delta_\uparrow E$ is measured from the Fe $\due$ states to the Mo
$\dut$ states. The position of the unoccupied Mo
$\dut$ states does not change.

This result contradicts Lu
\etal \cite{Lu2014ssc}, who observed instead a half-metallic state with a
spin transition from high spin to intermediate spin of the local
moment at the Fe ion. Hence, we present also calculations with the
FSM method including correlations corrections via GGA$+U$ (right side
of \Fref{fig:strain_magnetism}). The combination of FSM and GGA$+U$
allows to determine accurately the magnetic ground state of SFMO. For
compressive biaxial strain, we neither observe a spin transition nor a
half-metal to metal transition (\Fref{fig:strain_magnetism}h). The Fe
$\due$ states are now much further below the Fermi energy and
\SI{10}{\percent} compressive strain is not enough to shift them far
enough. We can only guess that the spin transition could be another
local minimum in the energy landscape of SFMO, while the FSM methods
ensures the correct magnetic ground state.

On the other hand, the spin transition in the 
tensile strain region at $\sim\SI{5}{\percent}$ 
(marked with a
vertical dashed line in \Fref{fig:strain_magnetism}h),
which is 
as well reported by
Lu \etal \cite{Lu2014ssc} for $\sim\SI{7}{\percent}$.
seems to be a stable magnetic phase transition.
While the exact biaxial strain value of the transition
may vary with the FSM treatment,
the ferrimagnetic state becomes
ferromagnetic (FIM-FM transition).
The magnetic moment of Fe ions decreases from 
$3.95\muB/\si{{f.u.}}$ to $3.50\muB/\si{{f.u.}}$,
while the Mo ions go into a low-spin state \cite{Lu2014ssc}
and their local moments align each other parallel
with the magnetic moments of the Fe ions
(\Fref{fig:strain_magnetism}e).

Hence, we can conclude that the
observed variations in the electronic and 
magnetic properties are rather small 
and not yet relevant 
in experimental conditions for SFMO 
(moderate strain values of \SIrange{-4}{4}{\percent})
\cite{Saloaro2016_10.1021/acsami.6b04132}. 
Nevertheless, we can observe an important effect
of biaxial strain for the formation of point defects
demonstrated in the next section.

\section{Point defects in SFMO with biaxial strain}

Using the supercell consisting of
$2\times2\times1$ tetragonal unit cells and 80 atoms (\Fref{fig:structure}c), we calculated
the defect formation energies of different point defects $D$ following
Nayak \etal 
\cite{Nayak2015prb} and considered in addition
a strain $\varepsilon$ dependence as $E_{\text{form}} (D, \varepsilon)$.
We briefly recall the 
technical details of determining the defect formation energy 
from the DFT total energies in the Supporting Information.

\begin{figure}
  \centering{
    \includegraphics[width=246pt]{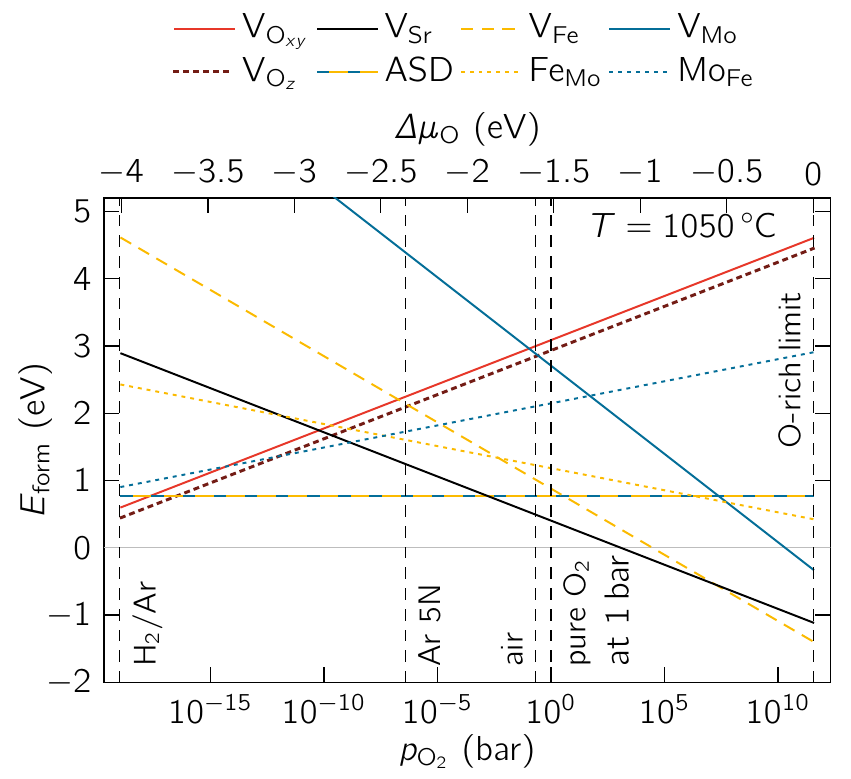}}
  \caption{(Color online) Defect formation energy of defects in unstrained SFMO ($\varepsilon=0$)
    for varying oxygen partial pressure at $T=\SI{1050}{\celsius}$.
    The relative chemical potential is shown at the upper axis.
    The legend is given above the figure.
  }
  \label{fig:thermodynamic_formation_energy}
\end{figure}

We considered besides common defect configurations such as oxygen vacancies
($\vO$) and antisite disorder (ASD) also
cation vacancies of Mo ($\vMo$), Fe ($\vFe$),
Sr ($\vSr$),
as well as non-stoichiometric disorder -- a Fe ion at a Mo sublattice
(Fe$_\mathrm{Mo}$) or a Mo ion at a Fe sublattice (Mo$_\mathrm{Fe}$).
We took also into account the two nonequivalent oxygen
positions, in-plane (O$_{xy}$) and out-of-plane (O$_{z}$) (\Fref{fig:relax}),
since the properties of the corresponding vacancies, $\vOxy$ and $\vOz$, respectively, might change considerably,
as shown for CaMnO$_3$ \cite{Aschauer2013prb}.
We note that the absolute values of formation energies are limited
to the choice of the chemical
potentials $\mu_i$, since the chemical potentials exactly suitable 
to the experimental growth conditions are unknown. 
Here, we chose the chemical potentials of the respective oxides
for all the chemical elements in SFMO (SrO, Fe$_2$O$_3$, MoO$_2$)
and the oxygen rich condition $\mu_\text{O}^\text{max}$
for the oxygen chemical potential \cite{Note1}.
In addition, the oxygen chemical potential can be varied 
as a parameter in order to simulate experiments at different oxygen partial 
pressure using thermodynamic considerations 
\cite{Note1, Reuter2001prb, Nayak2015prb}.
The results are straight lines with the slope $\propto\varDelta\mu_\text{O}=\mu_\text{O}-\mu^\mathrm{max}_\text{O}$
and $E_{\text{form}} (D, \varepsilon=0)$ at $\varDelta\mu_\text{O}=0$
(\Fref{fig:thermodynamic_formation_energy}). 
While the formation of  metal vacancies $\vFe$, 
$\vSr$ and  $\vMo$ is extremely
``easy'' due to the negative
$E_\text{form}(D,\varepsilon,\varDelta\mu_\text{O}=0)$, 
it is not surprising that both types of oxygen vacancies 
$\vOxy$, $\vOz$  (reddish lines) are less probable in the 
oxygen rich regime. With less oxygen partial pressure,
the formation energy of all metal vacancies increases and
their appearance gets more unlikely. Just at very low partial pressures
($\sim$ H$_2/$Ar atmosphere),
the formation energy of oxygen vacancies
becomes comparable with the one
of ASD. Hence, the relative probability of point defects 
are very sensitive to the exact experimental conditions.

We are well aware of other choices for reference
chemical potentials of multicomponent compounds using, e.g.,
the concept of constitutional defects developed by Hagen
and Finnis for ordered alloys \cite{Hagen1998_10.1080/01418619808223764},
which was applied for {\sfmo} \cite{Mishra2010cm}.
There the extreme case of Mo rich (Mo$_\mathrm{Fe}$)
or Fe rich Fe$_\mathrm{Mo}$ was used as references \cite{Mishra2010cm},
which makes a direct comparison complicated and we restrict ourselves
to the comparison of the electronic structure as discussed above.

At first, we verified for strain free SFMO that
GGA and FSM are applicable and we can avoid computational heavy calculations 
with, e.g., hybrid-functional methods,
which might raise problems for the calculations of defect
formation energies or need postprocessing as discussed in 
\cite{Nayak2015prb}. In particular, the choice of a specific $U$ for GGA$+U$ calculations
cannot be simply transferred to any of the reference systems used 
to determine the formation energy. Hence, we crosschecked again the
electronic structure including now the different defects in the 
supercell.
We observe that in agreement with other theoretical studies
\cite{Mishra2010cm,
  MunozGarcia2011_10.1021/cm201799c, Zhang2015jac_1}
only in the case of oxygen, iron, or strontium
vacancies, no additional states form in the spin up channel
at the Fermi energy and the DOS remains half-metallic
\cite{Note1}.

\begin{figure*}
  \centering{
    \includegraphics[width=512pt]{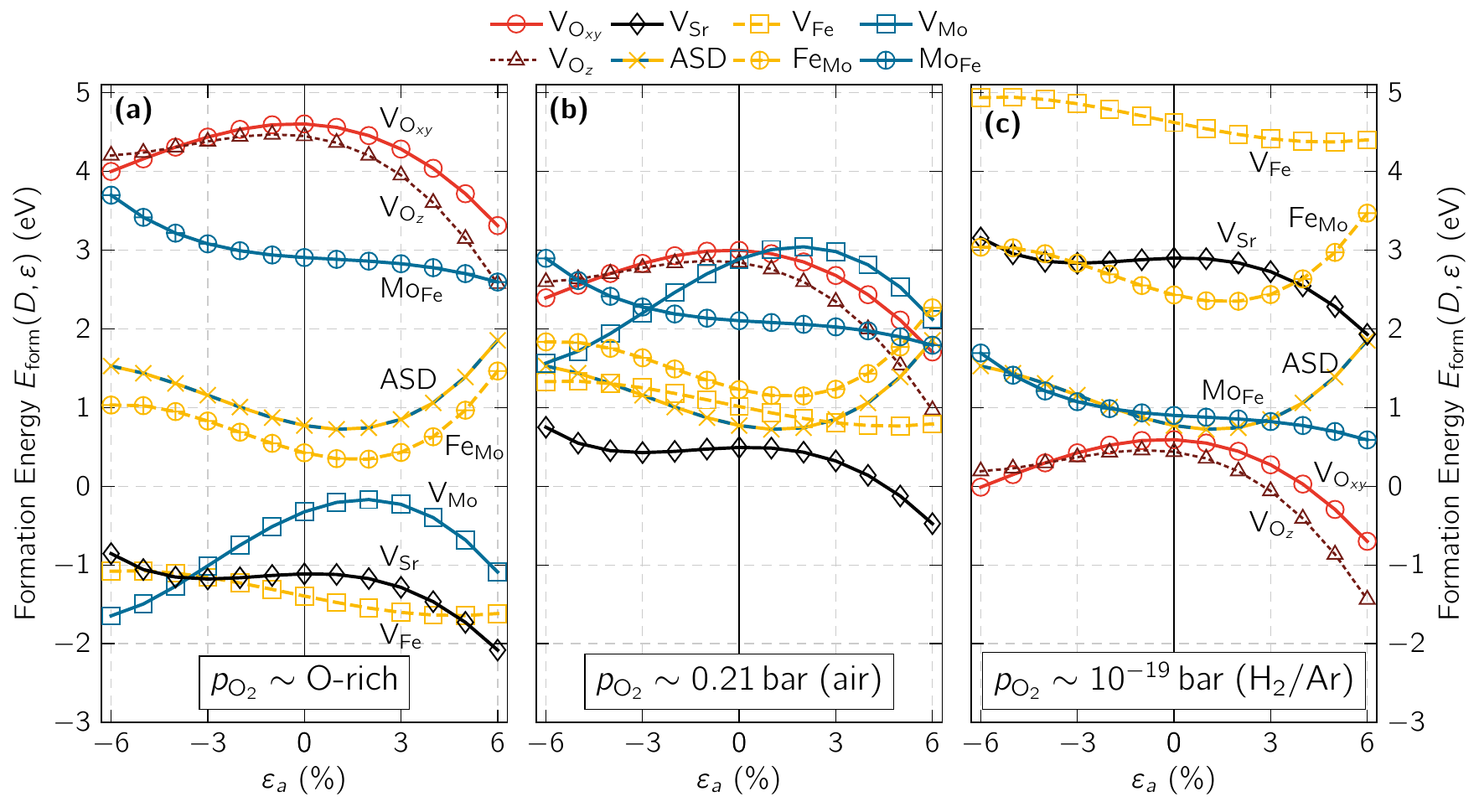}}
  \caption{(Color online) Formation energies of defects in dependence of biaxial strain and three
    different oxygen partial pressures. 
    The legend is given above the figure. $\vMo$ is out of range 
    in figure (c).}
  \label{fig:strain_defects}
\end{figure*}

The 
formation energies at biaxial strains follow then by
considering total energies of the relaxed supercells including defects,
scaled with respect to the defect-free, ground state total energy
for a series of moderate biaxial strains
(\SIrange{-6}{+6}{\percent}). We note that for compressive strains
$<\SI{-4}{\percent}$, the observed spin transition, discussed in the 
last section (\Fref[]{fig:strain_magnetism}), might cause also changes in the formation energy,
which would be hard to separate from purely strain mediated effects. 
Different oxygen partial pressure just scales all lines by an additive term 
depending on the respective chemical 
potentials with $\varDelta\mu_\text{O}$ or $p_{\text{O}_2}$ (\Fref[]{fig:thermodynamic_formation_energy}).
Hence, we obtained that the applied biaxial strain 
lowers for both  oxygen vacancy types -- 
$\vOxy$ and $\vOz$ -- the formation
energy and increases the probability of oxygen vacancies, 
with a more pronounced effect on the out-of-plane oxygen vacancy $\vOz$
(\Fref[]{fig:strain_defects}a).
This observation is in contrast with the results obtained for, 
e.g., CaMnO$_3$ \cite{Aschauer2013prb}, but we can also expect rather
different intrinsic strain relaxation mechanism in the largely distorted 
perovskite structure of CaMnO$_3$ and the less distorted SFMO structure.
The latter seems to compensate the compressive strain along the $z$ axis 
in the tensile regime with more oxygen vacancies in the out-of-plane direction,
while it becomes opposite for larger compressive strains
$\varepsilon\leq\SI{-5}{\percent}$ (maybe influenced by the spin transition).
Nevertheless, we can state that the oxygen vacancy formation is easier in 
biaxial strained films. This will be an advantage not only
for solid oxide fuel cells but also for the magnetic properties of SFMO, 
since we found recently that the Curie temperature can be strongly enhanced 
by oxygen vacancies in SFMO \cite{Hoffmann2015arxiv_SFMO}.

On the contrary, the formation energies of antisite defects increase 
for a large range of biaxial strain, e.g., 
(increase of $E_\text{form}(D)$) by \SI{+386}{\meV} and \SI{+78}{\meV} for compressive
or tensile strains $\varepsilon\pm\SI{3}{\percent}$ (\Fref{fig:strain_defects}).
It only becomes reduced by \SI{-48}{\meV} for small tensile strains in the range of
$0<\varepsilon<\SI{3}{\percent}$ and gets again larger for 
tensile strains $>\SI{2}{\percent}$.
That means that, in terms of applications,
biaxial strained films should be more ordered and better samples for spintronic applications
\cite{Saloaro2016_10.1021/acsami.6b04132}. 
This can be related to already made experimental observations 
on ASD. Various experiments investigated the $A$ site substitution
in double perovskites ($A_2BB'$O$_6$) and its influence to magnetization, 
spin polarization, ASD, etc. 
Chan \etal \cite{Chan2003_10.1021/cm020773h}
found a chemical pressure in SFMO, when substituting Sr with Ca, which 
caused an increase in magnetic moment but also an 
pronounced decrease in ASD. Hence, compression can, besides the chemical 
variations from Sr to Ca, reduce the number of anti-site defects.
Kahoul \etal \cite{Kahoul2008_10.1063/1.3043586} observed on the other hand
an increase in the lattice constant of SFMO, when doping Sr with La, and
this raised also the measured amount of ASD. Although the effects of chemical
substitution could overlay effects caused by a lattice constant 
variation, these experiments include hints of our 
theoretical observations. 

Considering the other defects, we found that
$\vSr$, $\vFe$ and Mo$_\mathrm{Fe}$
are less influenced by the biaxial strains \SIrange{-3}{3}{\percent}. 
Only $E_{\text{form}} (\vSr, \varepsilon)$
lowers considerably (\SI{-1}{\eV}) for larger tensile strains,
which then favors $\vSr$ over 
$\vFe$  (\Fref{fig:strain_defects}a).
Moreover, ASD can be considered as
a defect complex of the two non-stoichiometric defects
-- Mo$_\mathrm{Fe}$ and Fe$_\mathrm{Mo}$. The average
of their formation energies taking into account the 
oxygen chemical potential properly matches the one of
the antisite defects. 
The high formation energy of Mo$_\mathrm{Fe}$ is however
surprising, because we would expect substitution
of Fe with Mo much more likely, since 
SrMoO$_x$ is a typical impurity phase when growing
SFMO films \cite{Santiso2002sia_10.1002/sia.1435,
  Raittila2006jpcs_10.1016/j.jpcs.2006.03.009}.

This discrepancy is resolved when we consider lower oxygen partial pressures,
e.g., at ambient air pressure (\Fref{fig:strain_defects}b)
or almost vacuum (\Fref{fig:strain_defects}c).
Not only become $\vOxy$ more and $\vMo$ less likely,
also all formation energies shift to positive 
formation energies at $\varepsilon=0$. The latter means
that there are no spontaneous vacancies anymore and 
SFMO is indeed stable. Only at large tensile strains,
$E_\text{form}(\vSr)$ still gets below zero.
This means that biaxial strains can significantly alter the 
energy landscape --  the relative order of the defect formation energies
or their crossing points
(better visible in \Fref[]{fig:thermodynamic_formation_energy}). 
The low $E_\text{form}(\vSr)$ also means that $\vSr$ are very likely 
to appear in SFMO for a large range of experimental 
conditions. Therefore, Sr vacancies
could be more important than earlier expected and
might also deteriorate the magnetic properties of SFMO
\cite{Harnagea2015_10.1016/j.jssc.2014.11.017}. Also 
the formation of Mo oxide impurity phases is more likely
at ambient air partial pressure -- Fe vacancies form easily
and can be filled by Mo (Mo$_\text{Fe}$).

When going to lower partial pressures, antisite defects 
become the most likely defect (\Fsref{fig:thermodynamic_formation_energy} and 
\ref{fig:strain_defects}b) and only for partial pressures
smaller than \SI{e-16}{\bar} oxygen vacancies will be the most probable 
defect type (\Fref{fig:strain_defects}c). Then, non-stoichiometric 
Mo at an Fe site has an almost similar probability to form than an
antisite defect. This means that SrMoO$_x$
impurity phases are even more likely close to vacuum.

\section{Conclusions}

We showed that realistic biaxial strains will have a
pronounced effect on the point defect profile of oxide materials
-- here {\sfmo} -- via a strong influence on the defect stability
($E_{\text{form}} (D, \varepsilon)$ varies by
$\sim\SI{0.2}{\eV}$ per \SI{1}{\percent} strain).
The half-metallic ground state of defect-free SFMO will be 
preserved at moderate strains, but the relative 
order of formation energies for different defects might be 
altered, e.g., $\vFe$ vs. $\vSr$; or $\vOxy$ vs. $\vOz$. 

Most interestingly, the amount of
antisite defect formation can be reduced in compressively
strained SFMO films and oxygen vacancies will form 
much easier under both strains -- compressive and tensile.
Latter defects are crucial for solid oxide fuel cells
or can enhance magnetic properties in SFMO
\cite{Saloaro2016_10.1021/acsami.6b04132,Hoffmann2015arxiv_SFMO}. 

Taking into account also the oxygen partial pressure as an 
experimental parameter offers another degree of freedom
for the defect formation energies. 
Our numerical results at partial pressures matching
air or vacuum atmosphere
can be compared directly with experimental measurements. 
We found that $\vSr$ are very likely over a large range of
partial pressures, while oxygen vacancies become most likely only for 
very low partial pressures $<\SI{e-16}{\bar}$.
Hence, 
we have demonstrated the potential of targeted first-principles
calculations in designing strain-defect engineering
processes for the tuning of the properties of SFMO or
other oxides.
\section*{Acknowledgments}
This publication was funded by the German Research Foundation
within the Collaborative Research Centre 762 (projects A4 and B1).
\bibliographystyle{apsrev4-1}
\bibliography{journals,lib}



\section*{Supplementary material}

# Supplemental Material:
# Tuning the probability of defect formation via substrate strains in $Sr_2FeMoO_6$ films

Waheed A. Adeagbo,[1] Martin Hoffmann,[2] Arthur Ernst,[2,3] Wolfram Hergert,[1] Minnamari Saloaro,[4] Petriina Paturi,[4] and Kalevi Kokko[5,6]

[1]*Institute of Physics, Martin Luther University Halle-Wittenberg, Von-Seckendorff-Platz 1, 06120 Halle, Germany*[*]

[2]*Institute for Theoretical Physics, Johannes Kepler University Linz, Altenberger Straße 69, 4040 Linz, Austria*[†]

[3]*Max Planck Institute of Microstructure Physics, Weinberg 2, 06120 Halle, Germany*

[4]*Wihuri Physical Laboratory, Department of Physics and Astronomy, University of Turku, FI-20014 Turku, Finland*

[5]*Department of Physics and Astronomy, University of Turku, FIN-20014 Turku, Finland*

[6]*Turku University Centre for Materials and Surfaces (MatSurf), Turku, Finland*

---

[*] waheed.adeagbo@physik.uni-halle.de

[†] martin.hoffmann@jku.at

## A. VASP parameters

As exchange correlation functional, we used the generalized gradient approximation given in the Perdew-Burke-Ernzerhof (PBE) [1] form shortly noted as GGA. The projector augmented-wave (PAW) [2] method chosen for the treatment of core electrons yields a better ground state structure for oxide materials. The plane-wave expansion of the electronic wavefunctions was cut-off at the kinetic energy of 400 eV. The internal parameters and the lattice parameter $c$ were allowed to relax using the conjugate gradient algorithm until the forces were smaller then the tolerance value of 0.001 eV/Å. Atomic coordinates in tetragonal unit cell as a result of structural relaxation are given in Tab. I. We used a $\boldsymbol{k}$-point mesh of $8 \times 8 \times 6$ within the Monkhorst-Pack special $\boldsymbol{k}$-point scheme for the tetragonal cell and a $4 \times 4 \times 6$ $\boldsymbol{k}$-point mesh for the supercell to span the irreducible Brillouin zone [3] (see structures in Figure 1 of the main manuscript).

We calculated the elastic tensor for $Sr_2FeMoO_6$ (SFMO) from evaluations of the stress tensor generated by small strains [4].

## B. Density of States in defect-free $Sr_2FeMoO_6$

As shown already in many other theoretical papers, the density of states of SFMO reveals a half-metallic state (Fig. 1). However, the appearance of the band gap in the spin up channel in the numerical calculations is very sensitive to the particular numerical framework: One of the first calculations for example simply obtained half-metallic density of states [5]. On the other hand, the microstructure and the position of the oxygen ions with respect to the Fe and Mo ions seems to be crucial for the spin up band gap [6]. Lately, adding correlation corrections via DFT+$U$ approaches are named as the right tool in order to obtain half-metallicity [7–9]. Using the fixed spin moment approach and a generalized gradient approximation (GGA) exchange-correlation functional only, we obtained the half-metallic ground state with a total magnetic moment of $4\mu_B$/f.u.. The structural properties resemble very well experimental findings as discussed in the main manuscript.

Here, we want to emphasize on the density of states. The spin up band gap is for defect-free, unstrained SFMO 1.3 eV, measured between the unoccupied Mo $t_{2g}^\uparrow$ and the occupied Fe $e_g^\uparrow$ (Fig. 1). The position of the unoccupied Mo $t_{2g}^\uparrow$ explains also the "jump" of the spin

2

TABLE I. Structural relaxation of the tetragonal unit cell of unstrained SFMO ($\varepsilon = 0$) with FSM method and PBE functional. Lattice constants are $a = b = 5.5522\,\text{Å}$ and $c = 7.9013\,\text{Å}$. Internal positions are scaled with the respective lattice constants.

| site | $x$ | $y$ | $z$ |
|---|---|---|---|
| Sr | 0 | 1/2 | 1/4 |
| Sr | 0 | 1/2 | 3/4 |
| Sr | 1/2 | 0 | 1/4 |
| Sr | 1/2 | 0 | 3/4 |
| Fe | 0 | 0 | 1/2 |
| Fe | 1/2 | 1/2 | 0 |
| Mo | 0 | 0 | 0 |
| Mo | 1/2 | 1/2 | 1/2 |
| O | 0.21031 | 0.27868 | 0 |
| O | 0.78969 | 0.72132 | 0 |
| O | 0.72132 | 0.21031 | 0 |
| O | 0.27868 | 0.78969 | 0 |
| O | 0.71031 | 0.77868 | 1/2 |
| O | 0.28969 | 0.22132 | 1/2 |
| O | 0.22132 | 0.71030 | 1/2 |
| O | 0.77868 | 0.28969 | 1/2 |
| O | 0 | 0 | 0.24498 |
| O | 0 | 0 | 0.75502 |
| O | 1/2 | 1/2 | 0.74498 |
| O | 1/2 | 1/2 | 0.25502 |

up band gap described in the main manuscript. In the spin down channel, the Mo $t_{2g}^{\downarrow}$ and Fe $t_{2g}^{\downarrow}$ hybridize creating states at the Fermi energy. Applying an Hubbard $U$ of $4\,\text{eV}$ shifts the Fe $e_g^{\uparrow}$ downwards and increases the spin up band gap to $2.4\,\text{eV}$. This is of the same order as obtained by earlier calculations [8, 9], although the value is given as $\sim 1.9\,\text{eV}$ by Lu *et al.*

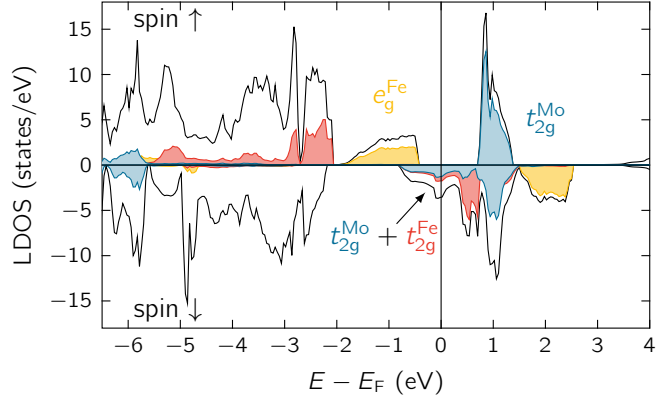

FIG. 1. Spin-resolved local density of states (LDOS) of unstrained and defect-free SFMO calculated with VASP.

citeLu2014ssc (the value seems to be counted only until the Fermi energy).

Although the spin up band gap in our case is rather small in comparison with other reports, the comparison with experimental values ($0.5\,\mathrm{eV}$, $1.3\,\mathrm{eV}$) shows a good agreement [10, 11]. The first one represents an optical measurement, while the second experimental band gap follows from photoemission spectroscopy. Of course, we have to consider also a thermal broadening of the electronic states with increasing temperature when comparing the size of the theoretical and experimental band gap.

### C. Electronic Properties of Point Defects in SFMO

Now, we consider also point defects in SFMO and calculate the band gap (Tab. II). As in the defect-free case, the Fe-$e_g^\uparrow$ states become more localized with GGA+$U$ and an already existing spin up band gap only widens by roughly $1\,\mathrm{eV}$ (Tab. II). Only for the non-stoichiometric defect Fe$_\mathrm{Mo}$ (additional Fe at a Mo site), a spin up band gap opens within the GGA+$U$, because the states of the additional Fe$_\mathrm{Mo}$ become more localized.

### D. Formation Energy

The formation energies provide a way to map the ground-state DFT calculations to the thermodynamic conditions accessible in experiment, and the relative comparison helps in judging the relative stability of various defects at a given environmental condition. We

4

TABLE II. Half metallic band gaps ($\Delta_\uparrow E$) of unstrained SFMO ($\varepsilon = 0$) including defects.

| configurations | $\Delta_\uparrow E$ (eV) | |
| --- | --- | --- |
| | GGA | GGA+$U$ |
| defect-free | 1.3 | 2.4 |
| ASD | 0.0 | 0.0 |
| $V_{O_{xy}}, V_{O_z}$ | 1.1 | 2.3 |
| $V_{Mo}$ | 0.0 | 0.0 |
| $V_{Fe}$ | 1.2 | 2.2 |
| $V_{Sr}$ | 1.3 | 2.2 |
| $Fe_{Mo}$ | 0.0 | 2.0 |
| $Mo_{Fe}$ | 0.0 | 0.0 |

calculated the formation energies in dependence of strain as [12]

$$E_{\text{form}}(D, \varepsilon) = E(D, \varepsilon, q = 0) - E_{\text{Host}}(\varepsilon) + \sum_i p_i n_i \mu_i. \quad (1)$$

Therein, $E(D, \varepsilon, q = 0)$ and $E_{\text{host}}(\varepsilon)$ are the calculated total energies of the supercell including defect ($D$), which is uncharged ($q = 0$, notation dropped in the following), and the host supercell of equal size under the influence of the strain $\varepsilon$, respectively. Equation (1) is balanced using the chemical potentials ($\mu_i$) of the elemental species ($i$). Since the energy dependence of the chemical potentials with strain was negligibly small compared with the difference between the total energies of strained bulk elements and their respective equilibrium structure (below 0.2 eV), we approximated the chemical potentials as strain independent Their variatione were not affecting the overall trend of the formation energies. In Equation (1), the number of defects is $n_i$ and the factor $p_i$ is either $-1$ or $+1$ for atoms added to or removed from the host supercell in the process of constructing the defect supercells [12, 13].

We took the chemical potentials used in (1) from the respective oxides of all the constituent metal elements. This means the total energy of SrO in the NaCl structure by

$$\mu_{\text{Sr}} + \mu_{\text{O}} = E(\text{SrO}). \quad (2)$$

and of $Fe_2O_3$ and $MoO_2$ ground state bulk phase for $\mu_{\text{Fe}}$ and $\mu_{\text{Mo}}$, respectively. The chemical potentials of oxygen $\mu_{\text{O}}$ was derived from half of the total energy of the oxygen molecule in

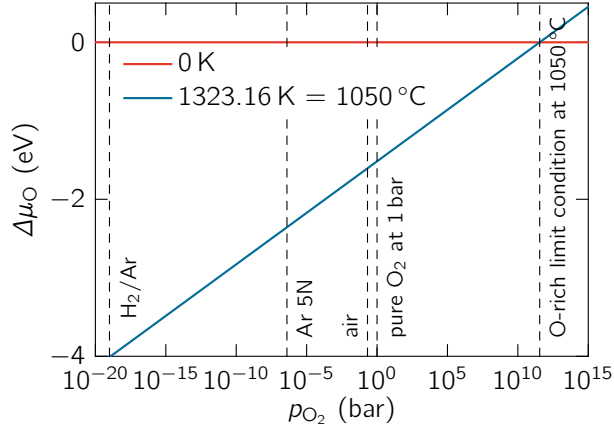

FIG. 2. Relation between oxygen partial pressure ($p_{O_2}$) and the relative chemical potential of oxygen ($\Delta\mu_O$) at $T = 0\,\text{K}$ and $1050\,°\text{C}$. The estimation is done by using (4) together with the absolute value for the oxygen chemical potential at O-rich condition set to zero [12].

the same volume as the supercell, i.e.,

$$\mu_O^{\max} = \frac{1}{2}E(O_2) = -4.916\,\text{eV}, \qquad (3)$$

which is considered as the oxygen rich limit.

However, the extremely rich case is not realized in experiments and the respective oxygen chemical potential will be reduced at experimental conditions. This can be seen from the thermodynamic relation between the oxygen chemical potential and the oxygen partial pressure ($p_{O_2}$) at a given temperature following the idea of Refs. [12, 14]

$$\mu(T, p_{O_2}) = \mu_O(T, p°) + \frac{1}{2}kT\ln\left(\frac{p_{O_2}}{p°}\right), \qquad (4)$$

where $p°$ is the standard pressure of $1\,\text{bar}$. The values of $\mu_O(T, p°)$ are calculated from the entropy and enthalpy of a oxygen molecule [14]

$$\mu_O(T, p°) = \frac{1}{2}[H(T, p°, O_2) - H(0\,\text{K}, p°, O_2)] - \frac{1}{2}T[S(T, p°, O_2) - S(0\,\text{K}, p°, O_2)]. \qquad (5)$$

Both quantities are listed in NIST-JANAF thermochemical tables [15]. In order to link the value of $\mu_O$ in Eq. (3) to the thermodynamic expression of Eq. (4), we choose a common reference $\mu_O(0\,\text{K}, p) = \mu_O^{\max}$ and defined the variation of the chemical potential as $\Delta\mu_O(T, p) = \mu_O(T, p) - \mu_O^{\max}$, which becomes $\Delta\mu_O(0\,\text{K}, p) = 0\,\text{eV}$ (Fig. 2). Finally, we assumed a temperature of $T = 1323.16\,\text{K} = 1050\,°\text{C}$, which is typically used when growing

6

epitaxial SFMO films [16–18]. We derived for $\mu_O(T, p^\circ) = -1.516\,27\,\text{eV}$ from the linear interpolation of equation (5) between 1300 K to 1400 K (Fig. 2).

With these equation, we could express the formation energy of the defects as a function of the oxygen partial pressure at different temperatures and obtained following equations

$$E_{\text{form}}(V_O, \varepsilon) = E(V_O, \varepsilon) - E_{\text{Host}}(\varepsilon) + (\Delta\mu_O + \mu_O^{\text{max}})\,, \tag{6}$$

$$E_{\text{form}}(V_{\text{Sr}}, \varepsilon) = E(V_{\text{Sr}}, \varepsilon) - E_{\text{Host}}(\varepsilon) + \left[E(\text{SrO}) - (\Delta\mu_O + \mu_O^{\text{max}})\right]\,, \tag{7}$$

$$E_{\text{form}}(V_{\text{Fe}}, \varepsilon) = E(V_{\text{Fe}}, \varepsilon) - E_{\text{Host}}(\varepsilon) + \left[\frac{1}{2}E(\text{Fe}_2\text{O}_3) - \frac{3}{2}(\Delta\mu_O + \mu_O^{\text{max}})\right]\,, \tag{8}$$

$$E_{\text{form}}(V_{\text{Mo}}, \varepsilon) = E(V_{\text{Mo}}, \varepsilon) - E_{\text{Host}}(\varepsilon) + \left[E(\text{MoO}_2) - 2(\Delta\mu_O + \mu_O^{\text{max}})\right]\,, \tag{9}$$

$$E_{\text{form}}(\text{Fe}_{\text{Mo}}, \varepsilon) = E(\text{Fe}_{\text{Mo}}, \varepsilon) - E_{\text{Host}}(\varepsilon) + \left[E(\text{MoO}_2) - 2(\Delta\mu_O + \mu_O^{\text{max}})\right] -$$
$$- \left[\frac{1}{2}E(\text{Fe}_2\text{O}_3) - \frac{3}{2}(\Delta\mu_O + \mu_O^{\text{max}})\right]\,, \tag{10}$$

$$= E(\text{Fe}_{\text{Mo}}, \varepsilon) - E_{\text{Host}}(\varepsilon) + E(\text{MoO}_2) - \frac{1}{2}E(\text{Fe}_2\text{O}_3) - \frac{1}{2}(\Delta\mu_O + \mu_O^{\text{max}})\,, \tag{11}$$

$$E_{\text{form}}(\text{Mo}_{\text{Fe}}, \varepsilon) = E(\text{Mo}_{\text{Fe}}, \varepsilon) - E_{\text{Host}}(\varepsilon) - \left[E(\text{MoO}_2) - 2(\Delta\mu_O + \mu_O^{\text{max}})\right] +$$
$$+ \left[\frac{1}{2}E(\text{Fe}_2\text{O}_3) - \frac{3}{2}(\Delta\mu_O + \mu_O^{\text{max}})\right]\,, \tag{12}$$

$$= E(\text{Mo}_{\text{Fe}}, \varepsilon) - E_{\text{Host}}(\varepsilon) - E(\text{MoO}_2) + \frac{1}{2}E(\text{Fe}_2\text{O}_3) + \frac{1}{2}(\Delta\mu_O + \mu_O^{\text{max}})\,, \tag{13}$$

while stoichiometric defects remain constant

$$E_{\text{form}}(\text{ASD}, \varepsilon) = E(\text{ASD}, \varepsilon) - E_{\text{Host}}(\varepsilon)\,. \tag{14}$$


\end{document}